\UseRawInputEncoding
\documentclass[%
 reprint,
superscriptaddress,
 amsmath,amssymb,
 aps, prb
]{revtex4-2}

\usepackage{graphicx}
\usepackage{dcolumn}
\usepackage{bm}
\usepackage{braket}
\usepackage[colorlinks=true, allcolors=blue]{hyperref}
\usepackage{cleveref}

\begin{document}
\bibliographystyle{apsrev4-2.bst}
\preprint{APS/123-QED}

\title{Si/SiO$_\text{2}$ MOSFET Reliability Physics: From Four-State Model to All-State Model}
\author{Xinjing Guo}
\affiliation{School of Microelectronics and Key Laboratory of Computational Physical Sciences (MOE), Fudan University, Shanghai 200433, China}
\author{Menglin Huang}
\email{menglinhuang@fudan.edu.cn}
\affiliation{School of Microelectronics and Key Laboratory of Computational Physical Sciences (MOE), Fudan University, Shanghai 200433, China}
\author{Shiyou Chen}
\email{chensy@fudan.edu.cn}
\affiliation{School of Microelectronics and Key Laboratory of Computational Physical Sciences (MOE), Fudan University, Shanghai 200433, China}
\affiliation{Shanghai Multi-scale Simulation Technology Co., Shanghai 201999, China}

\begin{abstract}
As implemented in the commercialized device modeling software, the four-state nonradiative multi-phonon model has attracted intensive attention in the past decade for describing the physics in negative bias temperature instability (NBTI) and other reliability issues of Si/SiO$_\text{2}$ MOSFET devices. It was proposed initially based on the assumption that the oxygen vacancy defects (V$_\text{O}$) in SiO$_\text{2}$ dielectric layer are bistable in the Si-dimer and back-projected structures during carrier capture and emission. Through high-throughput first-principles structural search, we found V$_\text{O}$ on non-equivalent O sites in amorphous SiO$_\text{2}$ can take 4 types of structural configurations in neutral state and 7 types of configurations in +1 charged state after capturing holes, which produce a wide range of charge-state transition levels for trapping holes. The finding contrasts the structural-bistability assumption and makes the four-state model invalid for most of O sites. To describe the reliability physics accurately, we propose an all-state model to consider all these structural configurations as well as all the carrier capture/emission transitions and thermal transitions between them. With the all-state model, we show that the V$_\text{O}$ defects play important roles in causing NBTI, which challenges the recent studies that discarded V$_\text{O}$ as a possible hole trap in NBTI. Our systematical calculations on the diversified V$_\text{O}$ properties and the all-state model provide the microscopic foundation for describing the reliability physics of MOSFETs and other transistors accurately.
\end{abstract}

\maketitle


\section{\label{sec:level1}Introduction}

As electronic devices are downscaled, reliability issues are becoming increasingly severe and must be considered for all types of field effect transistors (FETs), including planar metal-oxide-semiconductor FETs (MOSFETs), non-planar FinFETs, gate all around FETs (GAAFETs) and complementary FETs (CFETs) \cite{RN48,RN49,RN51,RN70,RN71}. The reliability issues in FETs are primarily manifested by the changes in the fundamental parameters that are used to characterize the performance of FETs \cite{RN72}, such as the threshold voltage (V$_\text{th}$), sub-threshold slope and on-current. The prominent reliability issues, such as bias temperature instability (BTI), random telegraph noise (RTN), and stress-induced leakage current (SILC), are believed to originate from the interaction of defects and impurities near Si/SiO$_\text{2}$ interface with carriers in the channel when a gate or drain-source voltage is applied \cite{RN11}. For instance, the negative BTI (NBTI) was attributed to the breaking of the interfacial Si-H bond under gate voltage, which induces the subsequent diffusion of H atom and generation of positively charged centers at the interface, and thus causes a negative shift of V$_\text{th}$ \cite{RN84}. This was referred as the reaction-diffusion (R-D) model, the most well-known model in the early studies of NBTI \cite{RN53}. However, the following NBTI recovery experiments showed a quick recovery of V$_\text{th}$, which cannot be explained by the R-D model that predicted a slow recovery behavior \cite{RN80,RN83}.  

\begin{figure*}[htbp]
\centering
\includegraphics[width=0.8\textwidth]{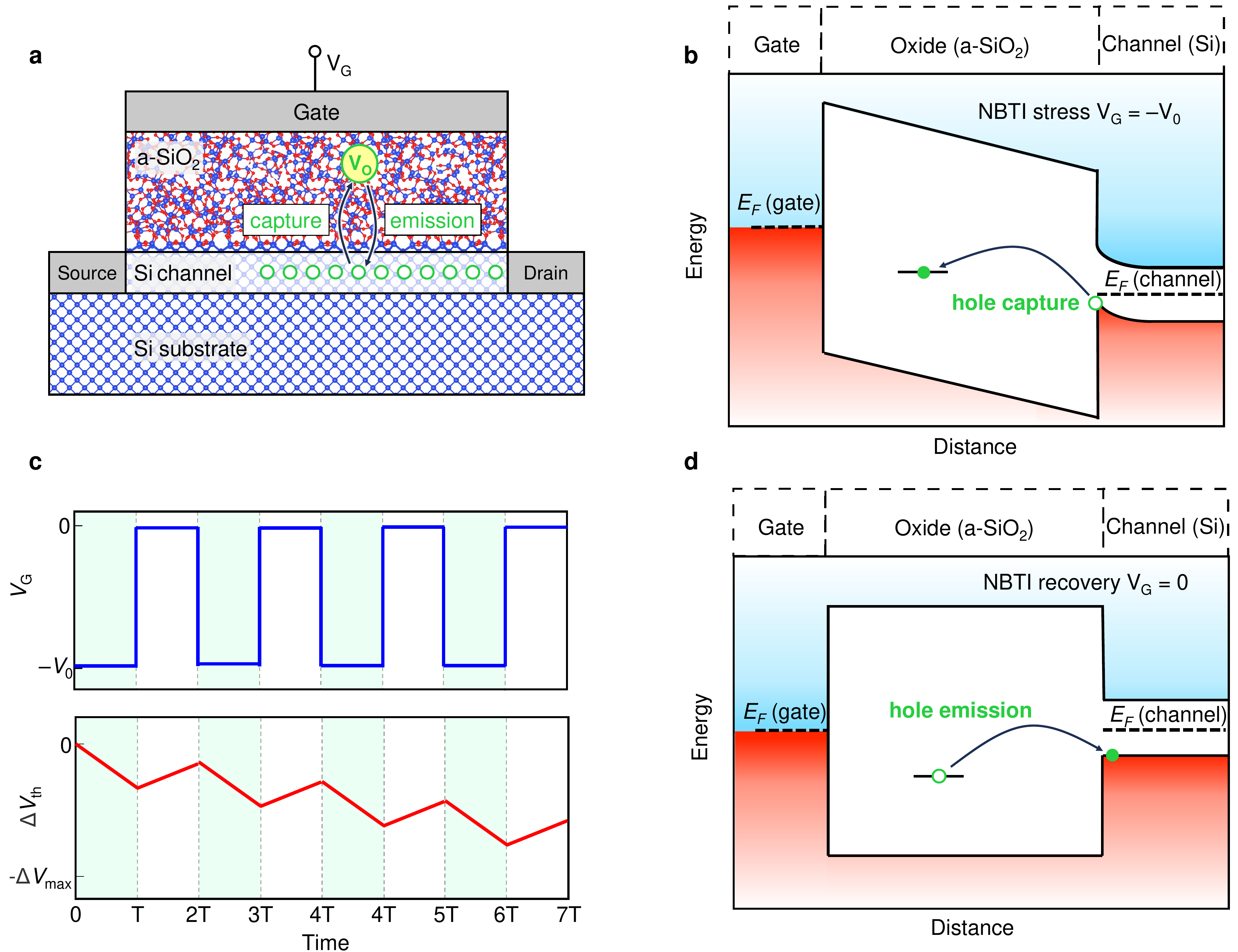}
\caption{\textbf{Hole capture/emission model of the NBTI degradation in Si/SiO$_\text{2}$ pMOSFET.} (a) Schematic plot of hole capture (emission) by the V$_\text{O}$ defects in SiO$_\text{2}$ layer from (to) the Si channel. (b) Band diagram of gate/oxide/channel layers under NBTI stress with V$_\text{G}$=$-$V$_\text{0}$. The negative V$_\text{G}$ produces a negative band slope and shifts the V$_\text{O}$ defect level up, driving V$_\text{O}$ to capture holes from Si channel. (c) Variation of $\Delta$V$_\text{th}$ during the repeated NBTI stress and recovery processes with a time period T. The gate voltage V$_\text{G}$ is biased at $-$V$_\text{0}$ under NBTI stress and is switched to 0 under NBTI recovery. (d) Band diagram under NBTI recovery with V$_\text{G}$=0. The V$_\text{O}$ defect level is shifted back and lower than the Si VBM level, driving V$_\text{O}$ to emit holes to Si. Here a flat band condition is supposed when V$_\text{G}$=0. }\label{Fig1}
\end{figure*}

To explain the quick NBTI recovery behavior, the hole capture/emission model (also known as trapping/detrapping model \cite{RN86}, or Kirton/Uren model \cite{RN85}), which attributes NBTI to the hole capture and emission by the defects in SiO$_\text{2}$, was proposed \cite{RN3,RN34,RN40}. As shown in \textbf{Figure~\ref{Fig1}a}, there may be pre-existing defects near Si/SiO$_\text{2}$ interface in pMOSFET, such as the oxygen vacancy (V$_\text{O}$), a predominant defect in SiO$_\text{2}$ \cite{RN35,RN37,RN4,RN5,RN6}. Under a negative gate voltage V$_\text{G}$ = $-$V$_\text{0}$ (NBTI stress), the energy band of SiO$_\text{2}$ dielectric layer as well as the V$_\text{O}$ defect level are shifted up relative to the valence band maximum (VBM) level of Si channel, so the V$_\text{O}$ defects can capture holes from the Si channel and transit into +1 charge state, as shown in \textbf{Figure~\ref{Fig1}b}. The cumulation of positive charges in the SiO$_\text{2}$ layer leads to a negative shift of V$_\text{th}$ (a negative $\Delta$V$_\text{th}$), as shown in \textbf{Figure~\ref{Fig1}c} where $\Delta$V$_\text{th}$ becomes more negative from t = 0 to t = T. When the gate voltage is removed (V$_\text{G}$ = 0, NBTI recovery), the V$_\text{O}$ defect level will be shifted down, and the holes will be emitted back to the Si channel, making V$_\text{O}$ become neutral, as shown in \textbf{Figure~\ref{Fig1}d}. The disappearance of positive charges on V$_\text{O}$ shifts the V$_\text{th}$ back, as shown in \textbf{Figure~\ref{Fig1}c} where $\Delta$V$_\text{th}$ becomes less negative from t = T to t = 2T. Since the shift of V$_\text{O}$ levels is instant once the gate voltage is applied or removed, the negative or backward shift of V$_\text{th}$ is also instant, so the hole capture/emission model can explain the quick NBTI recovery behavior successfully. 

Despite the success, the latter time-dependent defect spectroscopy experiments observed that the hole emission time constants of some defects depend on V$_\text{G}$, in agreement with the expectation according to the hole capture/emission model, while those of some other defects are independent of V$_\text{G}$ \cite{RN69,RN64}, which cannot be explained by the hole capture/emission model. To resolve this puzzling observation, Grasser et al. noticed the structural bi-stability of V$_\text{O}$ in SiO$_\text{2}$ and proposed the four-state model which attributes the different behaviors of hole emission time constants to the transitions between the four states of V$_\text{O}$ in SiO$_\text{2}$, including the ground state (Si-dimer structure) and metastable state (back-projected structure) of neutral V$_\text{O}$, as well as the ground state (back-projected structure) and metastable state (Si-dimer structure) of +1 charged V$_\text{O}$ after capturing a hole \cite{RN54}. After considering the ground-state and metastable structural configurations of both the neutral V$_\text{O}$ and +1 charged V$_\text{O}$, the four-state model can explain both the V$_\text{G}$-independent and V$_\text{G}$-dependent time constants \cite{RN11}. In recent years, the four-state model has become prevalent in the community and is now adopted for reliability modeling in many commercialized technology computer-aided design (TCAD) software such as Synopsys Sentaurus and Silvaco TCAD \cite{RN15,RN46}, e.g., for predicting the impact of defects on MOSFET \textit{I}-\textit{V} characteristics over a longer timescale \cite{RN14,RN13}.

As we can see, the structural bi-stability of V$_\text{O}$ defects is the foundation of the four-state model, however, whether the foundation is correct is still questionable. In crystalline $\alpha$-quartz SiO$_\text{2}$, V$_\text{O}$ indeed has such kind of structural bi-stability, i.e., the Si-dimer configuration is the ground-state structure and the back-projected configuration is the metastable structure of neutral V$_\text{O}$, while the back-projected configuration becomes the ground state and the Si-dimer configuration becomes metastable for +1 charged V$_\text{O}$ \cite{RN10,RN38}. In contrast, the crystalline $\alpha$-cristobalite SiO$_\text{2}$, a high-temperature phase \cite{RN39}, does not have such bi-stability, because first-principles calculations showed that the Si-dimer configuration is always the ground state of V$_\text{O}$, regardless of the charge state \cite{RN9}. This indicates the structures of V$_\text{O}$ defects are very sensitive to the crystal phase of SiO$_\text{2}$. The SiO$_\text{2}$ layer in Si/SiO$_\text{2}$ MOSFETs is usually the amorphous phase, in which all the O atomic sites become non-equivalent and the structure has a lower symmetry compared to the crystalline phase of SiO$_\text{2}$. Considering the sensitivity of V$_\text{O}$ defects to structures of SiO$_\text{2}$, we can expect the structural configurations of V$_\text{O}$ and their relative stability should be very complicated in a-SiO$_\text{2}$, which may not follow the structural bi-stability requirement of the four-state model. During the past two decades, the structural configurations of V$_\text{O}$ in a-SiO$_\text{2}$ have been studied through first-principles calculations \cite{RN6,RN18,RN7,RN8,RN24,RN43,RN47}. Besides the well-known Si-dimer and back-projected configurations, a series of new structural configurations of V$_\text{O}$ have been found \cite{RN6,RN7,RN8,RN24,RN43,RN47}, indicating the structural configurations in a-SiO$_\text{2}$ are indeed more complicated, which may go beyond the simple bi-stability. In 2022, Wilhelmer et al. \cite{RN19} performed high-throughput calculation studies on the possible configurations of V$_\text{O}$ in a 216-atom supercell and found that 95\% of V$_\text{O}$ defects take only two structural configurations, including the so-called unpuckered (Si-dimer) and puckered configurations. We noticed that they considered only two configurations (unpuckered and puckered) as the initial structures of high-throughput structural search and performed only local structural relaxation. However, the amorphous structure of SiO$_\text{2}$ has a low symmetry and thus a complicated potential energy surface with many local energy minimums, which may give rise to a large number of metastable structural configurations for the V$_\text{O}$ defects on different O sites \cite{RN58}. In order to find the ground-state (global energy minimum) structure and the important low-energy metastable structures of V$_\text{O}$, various initial structures (with different structural perturbations) should be included in the structural search and then the energy barriers between different structural configurations can be overcome. If only a very limited number of initial structures are used, the structural search is local, and important structural configurations may be missed. To evaluate whether the V$_\text{O}$ structural bi-stability foundation of four-state model is valid in a-SiO$_\text{2}$, finding all the important structural configurations is necessary. If there are other configurations with lower energy than the Si-dimer and back-projected configurations, the influences of these configurations cannot be neglected, which may make the structural bi-stability picture invalid in a-SiO$_\text{2}$, then the foundation of the widely-used four-state model becomes questionable and it may be inadequate for describing the MOSFET reliability physics. 

Although a series of new structural configurations of V$_\text{O}$ have been reported, whether these new configurations have high densities in a-SiO$_\text{2}$ is still an open question, and a global structural search of low-energy V$_\text{O}$ configurations in a-SiO$_\text{2}$ is also missing. Recently we developed the Defect and Dopant ab-initio Simulation Package (DASP) \cite{RN20}, which can perform nearly global structural search for defects through imposing random structural perturbations in the region around defects to produce various initial structures. Using DASP, we perform a high-throughput search of V$_\text{O}$ structures on different O sites in a-SiO$_\text{2}$. We find neutral V$_\text{O}$ can have four types of structural configurations: the Si-dimer, back-projected (including left- and right-back-projected), and double-back-projected configurations. For +1 charged V$_\text{O}$, in addition to the four configurations identified for neutral V$_\text{O}$, we also find three other configurations, including the left- and right-in-plane configurations and the twisted configuration, may also have low energy and thus high density. Depending on the O sites, the ground-state structures of +1 charged V$_\text{O}$ can vary among the seven configurations, so their roles in the hole capture/emission should be important and non-negligible. The increased number of low-energy structural configurations makes the structural bi-stability foundation of the four-state model invalid. In order to describe the microscopic mechanisms of NBTI and other reliability issues more accurately, we propose an all-state model that considers all the identified structural configurations of V$_\text{O}$ in a-SiO$_\text{2}$ as well as the multi-phonon non-radiative hole capture/emission transitions and the thermal transitions between these configurations. With the all-state model, we show that the various configurations of V$_\text{O}$ defects in a-SiO$_\text{2}$ produce a large number of transition levels with a wide energy range, so they can act as hole trapping/detrapping centers and play important roles in NBTI, which challenges the recent studies that discarded V$_\text{O}$ as hole trapping/detrapping centers in NBTI and re-establishes the importance of V$_\text{O}$ defects. These results demonstrate the necessity of considering all the states (structural configurations and charge states) of the defects in amorphous oxide dielectric layers that may act as the origin of BTI, RTN, SILC, or other reliability issues, so the all-state model should be general for describing the physics of all these reliability issues.

\section{\label{sec:level1}Results}

\subsection{\label{sec:level2}Nonequivalent V$_\text{O}$ sites in a-SiO$_\text{2}$}

\begin{figure*}[htbp]
\centering
\includegraphics[width=0.8\textwidth]{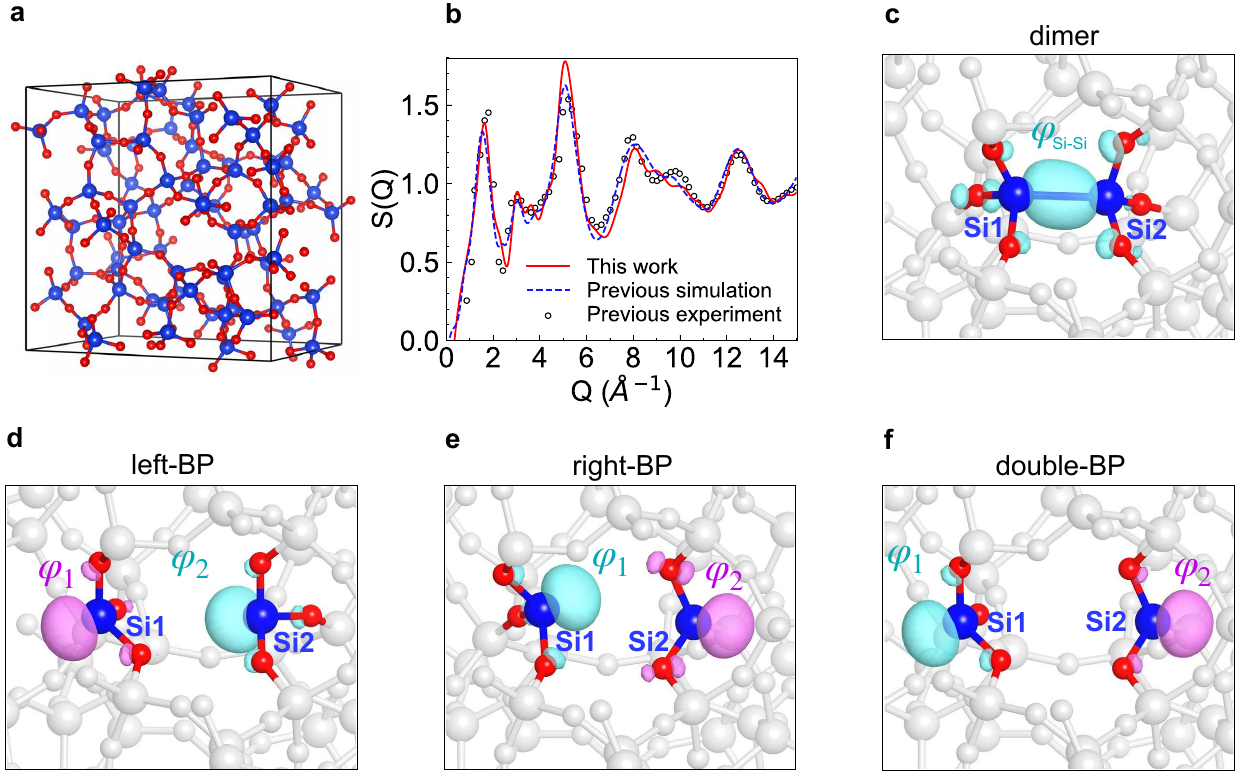}
\caption{\textbf{Structural model of a-SiO$_\text{2}$ and neutral V$_\text{O}$ defects.} (a) Supercell model generated by bond-switching Monte Carlo method for mimicking amorphous SiO$_\text{2}$. (b) Structure factor of the supercell model, in agreement with theoretical results reported in Ref. \cite{RN19} and experimental results reported in Ref. \cite{RN22}. (c-f) Local structures of V$_\text{O}$ defects with Si-dimer, left-back-projected (left-BP), right-back-projected (right-BP), and double-back-projected (double-BP) configurations, as well as the norm-squared wavefunctions of the defect states. For Si-dimer configuration in (c), $\varphi_\text{Si-Si}$ is the bonding state of the $\sigma$ bond. In (d-f), $\varphi_\text{1}$ is the defect state localized on Si1 atom, while $\varphi_\text{2}$ is localized on Si2 atom. }\label{Fig2}
\end{figure*}

To study the V$_\text{O}$ defects in amorphous SiO$_\text{2}$, we adopt the bond-switching Monte Carlo method \cite{RN21} to generate a supercell model that replicates the amorphous structure of SiO$_\text{2}$. A 216-atom supercell of $\alpha$-quartz SiO$_\text{2}$ phase is used as the initial structure of the Monte Carlo simulation, and the generated supercell model is shown in \textbf{Figure~\ref{Fig2}a}. As shown in \textbf{Figure~\ref{Fig2}b}, its calculated structure factor \cite{RN44} is in good agreement with that measured in experiment \cite{RN22} and that reported in the previous simulation study of a-SiO$_\text{2}$ \cite{RN19}. Details about the MC method and the supercell size convergence test are shown in Methods Section and Supplementary Material. 

In the supercell model, there are 144 non-equivalent O sites. Due to the complicated atomic coordination and varying bond lengths around O on different sites, the properties of V$_\text{O}$ formed on these O sites may differ. Therefore, a comprehensive understanding of V$_\text{O}$ defect properties requires an investigation of V$_\text{O}$ on all these non-equivalent sites. In SiO$_\text{2}$, V$_\text{O}$ usually acts as a donor defect. In neutral state, the defect level of the un-ionized V$_\text{O}$ is fully occupied by electrons. If a hole is captured by V$_\text{O}$, rendering the defect level unoccupied, V$_\text{O}$ becomes ionized and stays in +1 charge state. In the following, we will start from neutral V$_\text{O}$ in a-SiO$_\text{2}$, searching for its low-energy structural configurations on different sites and exploring their electronic structures, and then perform these studies for +1 charged V$_\text{O}$.  

\subsection{\label{sec:level2}Possible structures of neutral V$_\text{O}$}

In previous studies, the structural configurations of V$_\text{O}$ are usually obtained by removing one oxygen atom from the a-SiO$_\text{2}$ structure and then performing local structural relaxations, which typically finds the Si-dimer configuration (sometimes referred as unpuckered configuration) \cite{RN19}. In order to search for other structural configurations of V$_\text{O}$ in crystalline and amorphous SiO$_\text{2}$, people also tried to move the neighboring Si atom through the plane of its three adjacent O atoms, and then perform structural relaxation, which results in the so-called back-projected configuration \cite{RN7,RN24} (sometimes referred as puckered configuration \cite{RN18,RN19}). Following these studies, we also find the Si-dimer configuration if no structural perturbation is added, and find the back-projected configuration if the Si back-projected perturbation is added, in good agreement with previous studies \cite{RN7,RN24}. Besides the Si back-projected perturbation, there are also other types of structural perturbations that may give rise to new structures after structural relaxation. Since the amorphous structure has a low symmetry and complicated potential energy surface, these new structural configurations of V$_\text{O}$ may have low energies. Therefore, we should add many types of structural perturbations and perform a global structural search of V$_\text{O}$, in order to identify the ground-state and low-energy metastable configurations that play important roles in Si/a-SiO$_\text{2}$ MOSFET reliability physics. 

We perform the global search through three steps. Firstly, we choose one O site in a-SiO$_\text{2}$ structure and generate 40 different random structural perturbations for the V$_\text{O}$ on that site using DASP software \cite{RN20}. To ensure the 40 perturbations are different from each other, we adopt the quantitative descriptor of structural difference $\Delta$Q (the definition of $\Delta$Q is given in Method Section), and require $\Delta$Q $>$ 1 amu$^{1/2}$ \AA \ for all perturbations. Then we carry out relaxations for these 40 configurations, and identify four types of V$_\text{O}$ configurations, namely Si-dimer, left-back-projected, right-back-projected, and double-back-projected configurations, which will be described later in detail. Secondly, in order to reflect the structural characteristics of different O sites in a-SiO$_\text{2}$, we also choose four other O sites and generate 40 perturbed structures for V$_\text{O}$ on each site, and then perform relaxations for these 160 configurations. The result shows all the perturbed configurations are relaxed to the four identified configurations. Thirdly, for the remaining 139 O sites among the 144 O sites in the 216-atom a-SiO$_\text{2}$ supercell, we suppose that V$_\text{O}$ on these sites also take the four identified configurations, and add the corresponding structural perturbations for each site, which gives rise to a total of 139×4=556 configurations. Then we carry out structural relaxation for these V$_\text{O}$ configurations, and find the relaxed structures fall within the scope of the four configurations. In the following, we will introduce the configurations of V$_\text{O}$ after the relaxation.

\begin{figure}[htbp]
\centering
\includegraphics[width=0.5\textwidth]{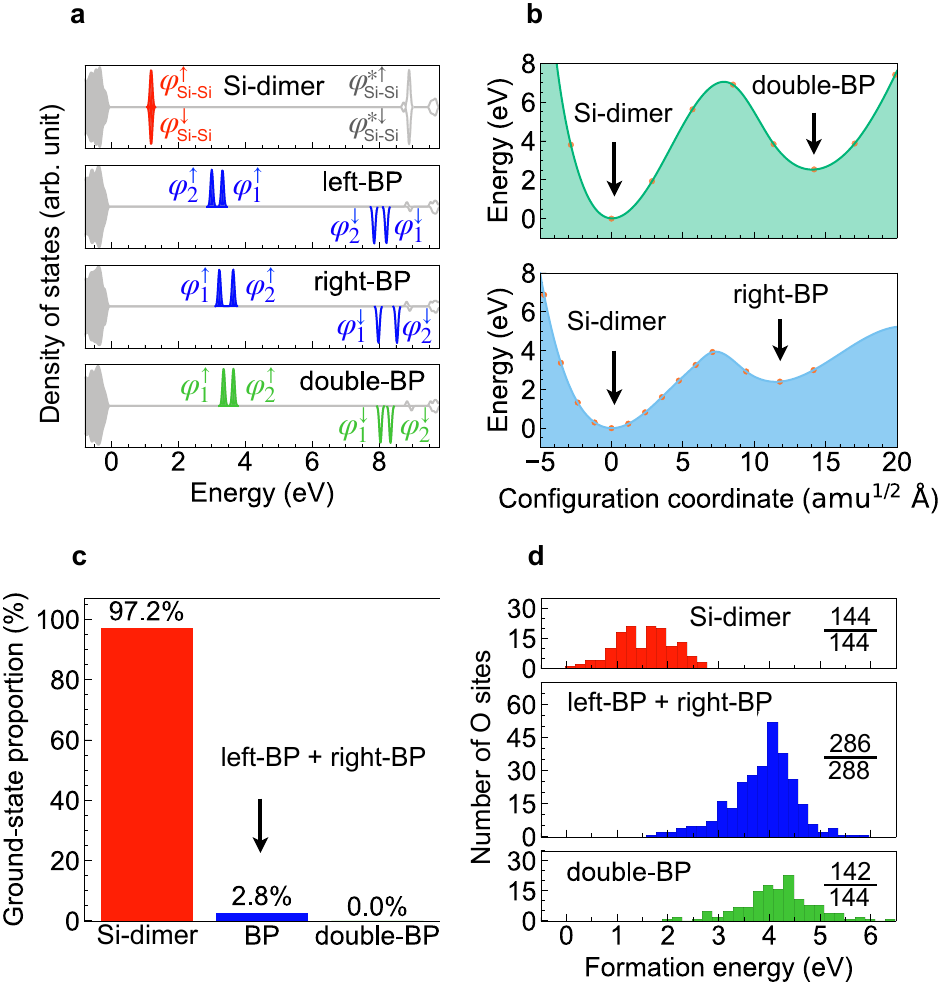}
\caption{\textbf{Electronic states and energetic stability of neutral V$_\text{O}$ in the four identified structural configurations.} (a) Density of states of V$_\text{O}$ in these configurations. One O site is taken as an example. The symbols for defect states are consistent with those in Figure~\ref{Fig1}. (b) Potential energy surfaces of transitions from Si-dimer to double-back-projected and to right-back-projected configurations. (c) Proportion of different configurations acting as the ground state of V$_\text{O}$ on all 144 O sites. (d) Formation energy distribution of different configurations. The distribution is calculated by counting the number of O sites on which the V$_\text{O}$ defects in a configuration have formation energies in the 0.2 eV (bin width) energy range. The distribution of back-projected configurations is the sum of the left- and right-back-projected configurations. The ratio labeled in each panel shows the proportion of O site where the intentionally perturbed configuration on this O site can be maintained after structural relaxation.}\label{Fig3}
\end{figure}

The first one is the Si-dimer configuration. When an O atom is removed from the crystal, the two adjacent Si1 and Si2 atoms may move closer, forming a Si p–Si p $\sigma$ bond. The wavefunction of the low-energy bonding state is shown in \textbf{Figure~\ref{Fig2}c}, and it introduces two occupied levels (spin-up and spin-down) in the band gap of SiO$_\text{2}$, as shown in the density of states in \textbf{Figure~\ref{Fig3}a}. Correspondingly, there are also two antibonding states whose wavefunctions are plotted in Figure S10, and their energy levels are unoccupied and slightly higher than the CBM (\textbf{Figure~\ref{Fig3}a}). Since spin-up $\varphi_\text{Si-Si}^\uparrow$ and spin-down $\varphi_\text{Si-Si}^\downarrow$ levels are degenerate and both occupied, the calculated total spin S = 0, and thus this defect configuration is non-paramagnetic, making neutral V$_\text{O}$ inactive in electron paramagnetic resonance (EPR) \cite{RN57}. On 140 of all 144 O sites, the Si-dimer configuration is the ground state (lowest energy) among all V$_\text{O}$ configurations. On the other 4 O sites, it is metastable and does not transit to other configurations during local structural relaxation.

Compared to the Si-dimer configuration, the back-projected configuration has a large structural reorganization. For example, the left-back-projected configuration is formed if the left-side Si1 atom moves backward through the plane of its three O neighbors and stays on the left side of the O plane. Because of the movement of Si1 atom, the Si-Si bond in the Si-dimer configuration is broken, producing two dangling bonds at the Si1 and Si2 atoms and two dangling bond states ($\varphi_\text{1}$ and $\varphi_\text{2}$), as shown in \textbf{Figure~\ref{Fig2}d}. Each dangling bond has one electron. The two electrons occupy the spin-up $\varphi_\text{1}^\uparrow$ and $\varphi_\text{2}^\uparrow$ levels, while the spin-down $\varphi_\text{1}^\downarrow$ and $\varphi_\text{2}^\downarrow$ levels are unoccupied, giving a total spin S = 1, as shown by the density of states in \textbf{Figure~\ref{Fig3}a}. There is an exchange splitting around 5 eV. Similar to Si1 atom, the Si2 atom on the right side can also move backward to the right side of the O plane, forming the right-back-projected configuration, as plotted in \textbf{Figure~\ref{Fig2}e}. It is important to note that the Si1 and Si2 atoms are nonequivalent and asymmetric due to the low symmetry of amorphous structure, which is different from the case in crystalline $\alpha$-quartz SiO$_\text{2}$ where the two Si atoms near V$_\text{O}$ are equivalent. There is an energy barrier between the Si-dimer configuration and (left or right) back-projected configurations, as shown in \textbf{Figure~\ref{Fig3}b}. As a result, when an O atom is removed and one V$_\text{O}$ is formed, the structure relaxes directly to the Si-dimer configuration on all 144 O sites. Only when the structural perturbations are added to overcome the barrier, can the left-back-projected and the right-back-projected configurations be found during structural relaxation. In this way, the left- and right-back-projected configurations can be both found on 142 of the 144 O sites, while on the remaining 2 O sites, only one of the left- and right-back-projected configurations can be maintained during structural relaxation. The left- or right-back-projected configuration has the lowest energy and serves as the ground state on 4 O sites, where the Si-dimer configuration has a higher energy.

If both the Si1 and Si2 atoms undergo a large structural reorganization and move backward through the O planes on the left and right side, respectively, the double-back-projected configuration is formed. Similar to the back-projected configuration, the dangling bonds at the Si1 and Si2 atoms produce two states $\varphi_\text{1}$ and $\varphi_\text{2}$, as shown by the contour in \textbf{Figure~\ref{Fig2}f}. Their corresponding defect levels are shown in the density of states in \textbf{Figure~\ref{Fig3}a}. If both the left- and right-back-projected configurations exist on one O site, our calculations show that the double-back-projected configuration also exists on this O site. Therefore, the double-back-projected configuration exists on 142 O sites. On the 2 O sites where only one of the left- and right-back-projected configurations exists, the double-back-projected configuration is unstable during structural relaxation. The calculations on all 144 O sites show that the double-back-projected configuration does not act as the ground state on any O site.

The electronic structure of double-back-projected configuration is similar to those of left- and right-back-projected configurations, i.e., their $\varphi_\text{1}$ and $\varphi_\text{2}$ defect states all originate from the dangling bonds of Si1 and Si2 atoms, and the spin-up $\varphi_\text{1}^\uparrow$ and $\varphi_\text{2}^\uparrow$ levels are occupied by the two electrons, giving a total spin S = 1, as shown in \textbf{Figure~\ref{Fig3}a}. Interestingly, the electron paramagnetic resonance (EPR) has unveiled a metastable V$_\text{O}$ related center with a total spin S = 1 in a-SiO$_\text{2}$, where the two unpaired spins were localized at two Si dangling bonds and this signal was removed after a 10-min anneal at T $\geq$ 200 $^{\circ}C$ \cite{RN56}. This observation cannot be explained by the Si-dimer configuration of neutral V$_\text{O}$ which has a total spin S = 0, but it is in accordance with our calculated spin and the defect charge densities of back-projected and double-back-projected configurations, indicating that the defect observed by EPR can probably be the back-projected or double-back-projected configurations of neutral V$_\text{O}$. It should be noted that our calculations also show that the back-projected and double-back-projected configurations on some O sites may have higher-energy electronic configurations with S = 0: (i) one electron localized at  Si1  atom occupies the spin-up $\varphi_\text{1}^\uparrow$ level, while the other electron localized at Si2 atom occupies the spin-down $\varphi_\text{2}^\downarrow$ level, giving a total spin S = 0; (ii) two electrons are both localized at  Si1  atom and occupy $\varphi_\text{1}^\uparrow$ and $\varphi_\text{1}^\downarrow$ levels, while the back-projected Si2 atom forms a covalent bond with another O atom and passivates its dangling bond, giving a total spin S = 0 as well. The defect charge densities and density of states of these higher-energy electronic configurations are analyzed in Supplementary Material.

The Si back-projected structural reorganization from Si-dimer to back-projected configuration does not occur spontaneously for neutral V$_\text{O}$, because the potential energy surfaces of the transition have a large energy barrier, as demonstrated in \textbf{Figure~\ref{Fig3}b} for V$_\text{O}$ on one O site. Therefore, the successful finding of back-projected or double-back-projected configurations during local structural relaxation requires a sufficient structural perturbation to overcome the transition barrier. That confirms the necessity of generating 40 different random structural perturbations in our structural search.

\textbf{Figures \ref{Fig3}c} and \textbf{\ref{Fig3}d} compare the energetic stability of different neutral V$_\text{O}$ configurations. As shown in \textbf{Figure~\ref{Fig3}c}, 97.2\% of all O sites take the Si-dimer configuration as their ground-state structures of V$_\text{O}$, while the back-projected and double-back-projected configurations only account for 2.8\% and 0\%. This trend is also obvious in the distribution of their formation energies plotted in \textbf{Figure~\ref{Fig3}d}. The energy distribution of the Si-dimer configuration is generally much lower than that of the back-projected and double-back-projected configurations, and the energy of double-back-projected configuration is slightly higher than that of back-projected configuration. Such an order in formation energies is consistent with the energies of the occupied levels of the V$_\text{O}$ defect states, as shown by the density of states in \textbf{Figure~\ref{Fig3}a}. On the O site, the occupied $\varphi_\text{Si-Si}^\uparrow$ level of Si-dimer configuration is 1.81 eV and 2.13 eV lower than the occupied $\varphi_1^\uparrow$ and $\varphi_2^\uparrow$ levels of left-back-projected configuration respectively, while the formation energy of Si-dimer configuration is 1.67 eV lower than that of left-back-projected configuration, so their formation energy difference is contributed mainly by the energy level difference of occupied defect states. According to the stability analysis, the back-projected and double-back-projected configurations act mainly as metastable structures of neutral V$_\text{O}$ and their energies are generally higher, so most of the neutral V$_\text{O}$ defects in a-SiO$_\text{2}$ should take the most stable Si-dimer configuration. This also explains the EPR observation that the signal of V$_\text{O}$ related center with a spin S = 1 was removed after a 10-min anneal at T $\geq$ 200 $^{\circ}C$ \cite{RN56}, because the higher-energy back-projected and double-back-projected configurations with S = 1 tend to transit into the more stable Si-dimer configuration with S = 0.

\subsection{\label{sec:level2}Possible structures of +1 charged V$_\text{O}$}

In the aforementioned search, we identified four types of configurations for neutral V$_\text{O}$. Under NBTI stress, neutral V$_\text{O}$ can capture a hole from Si VBM, becoming +1 charged. An important question arises as to whether these +1 charged V$_\text{O}$ also exhibit the same four configurations as their neutral counterparts. Given that structural perturbations are required for overcoming the transition barriers between different configurations of neutral V$_\text{O}$, it is necessary to add similar perturbations for exploring the configurations of +1 charged V$_\text{O}$. Therefore, we choose 5 O sites and add 40 random perturbations to the +1 charged V$_\text{O}$ structure on each O site, generating a total of 200 configurations. Structural relaxation shows all these configurations converge into 7 different types of configurations. Among them, four types (Si-dimer, left-back-projected, right-back-projected, and double-back-projected configurations) have already been found in the search of neutral V$_\text{O}$ configurations, whereas three new types of configurations (namely left-in-plane, right-in-plane, and twisted configurations) are only found for +1 charged V$_\text{O}$. Now we will discuss their atomic structures and electronic states in order. 

\par
\begin{figure*}[htbp]
\centering
\includegraphics[width=0.82\textwidth]{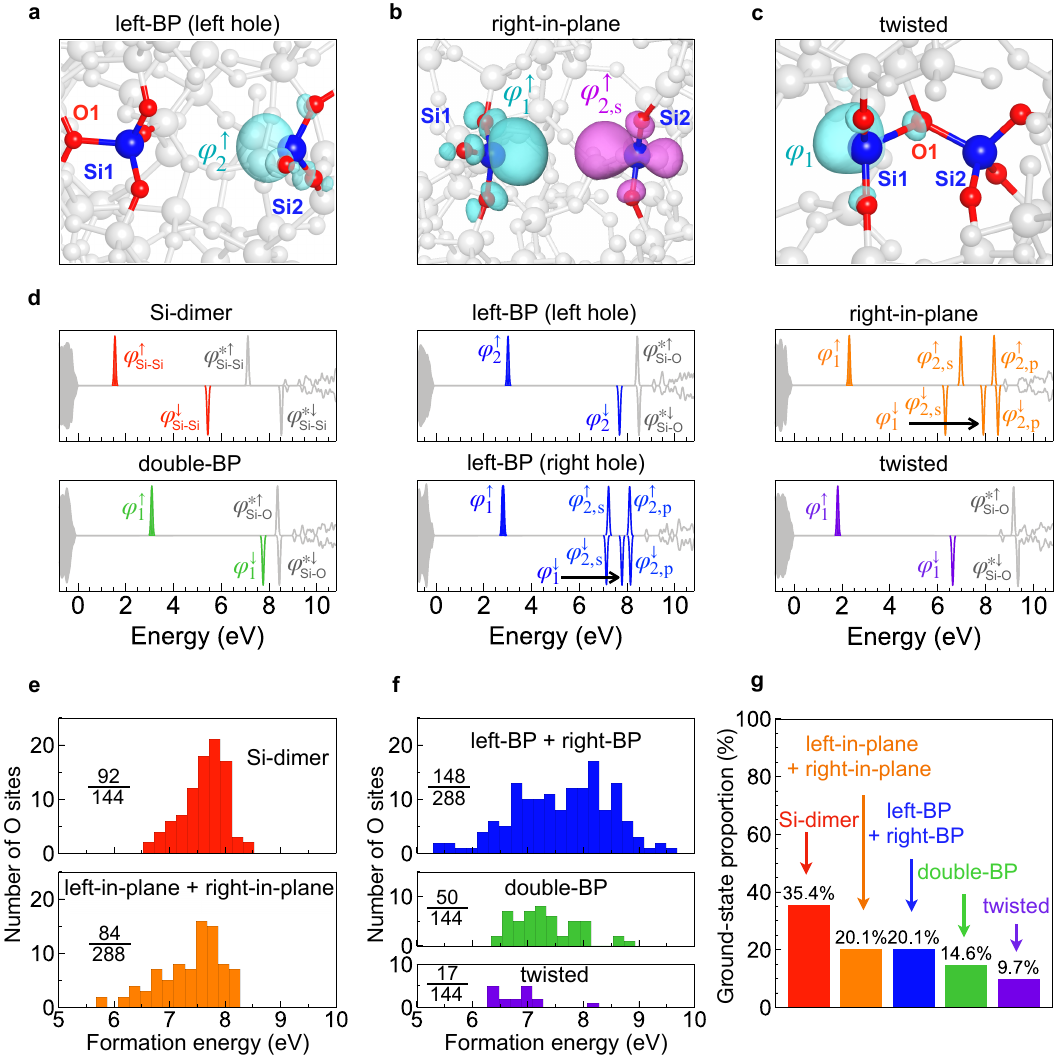}
\caption{\textbf{Structures, electronic states and energetic stability of +1 charged V$_\text{O}$.} (a-c) Local geometric structures of +1 charged V$_\text{O}$ with left-back-projected (left hole), right-in-plane, and twisted configurations. The contour shows the norm-squared wavefunction of the corresponding defect states. (d) Electronic density of states for Si-dimer, left-back-projected (left hole), right-in-plane, double-back-projected, left-back-projected (right hole), and twisted configurations. (e, f) Formation energy distribution of +1 charged V$_\text{O}$ in Si-dimer, in-plane, back-projected, double-back-projected, and twisted configurations. The distribution is calculated by counting the number of O sites on which the V$_\text{O}$ defects in the configuration have formation energies in the 0.2 eV (bin width) energy range. Fermi level is assumed at the middle of the band gap. The ratio labeled in each panel shows the proportion of O site where the intentionally perturbed configuration on this O site can be maintained after structural relaxation. (g) Proportion of different V$_\text{O}$ configurations acting as the ground state on all 144 O sites.}\label{Fig4}
\end{figure*}

 For the V$_\text{O}$ Si-dimer configuration, its atomic structure in +1 charge state is similar to its neutral counterpart (\textbf{Figure~\ref{Fig2}c}), but has a longer Si-Si bond and thus a weaker $\sigma$ bond, as shown in Figure S9, Figure S11 and Figure S14. The longer Si-Si bond is caused by the Coulomb repulsion between Si cations and the trapped hole by V$_\text{O}$. According to the calculated density of states in \textbf{Figure~\ref{Fig4}d}, the electron is removed from the spin-down bonding state ($\varphi_\text{Si-Si}^\downarrow$), making it unoccupied. The exchange splitting enlarges the energy separation between occupied $\varphi_\text{Si-Si}^\uparrow$ and unoccupied $\varphi_\text{Si-Si}^\downarrow$ states. Due to the weakened hybridization, the bonding states ($\varphi_\text{Si-Si}^\uparrow$ and $\varphi_\text{Si-Si}^\downarrow$) are shifted up and the antibonding states ($\varphi_\text{Si-Si}^{*\uparrow}$ and $\varphi_\text{Si-Si}^{*\downarrow}$) are shifted down, compared to those in neutral state shown in \textbf{Figure~\ref{Fig3}a}. 

For the left-back-projected configuration (similarly for the right-back-projected configuration) of +1 charged V$_\text{O}$, the captured hole can be trapped to the left Si1 or right Si2 atom, giving rise to two sub-configurations: left hole and right hole. \textbf{(i)} For the left hole case, the captured hole is trapped to the left Si1 atom, meaning that the Si1 dangling-bond electron occupying the $\varphi_\text{1}^{\uparrow}$ level of the neutral left-back-projected configuration (\textbf{Figure~\ref{Fig2}d}) is removed from the V$_\text{O}$ site, so the Si1 atom may displace towards the left side and bonds with another O atom (O1), becoming fourfold coordinated (\textbf{Figure~\ref{Fig4}a}). For +1 charged V$_\text{O}$ on some O sites, the left Si1 atom trapping the hole can also bond with the O1 atom and another Si atom (Si3) simultaneously, becoming fivefold coordinated (Figure S12). The left-back-projected (left hole) configuration has been mentioned in literature \cite{RN7}. Eventually, the passivation of the original dangling bond of Si1 leads to the elimination of the defect-state ($\varphi_\text{1}^{\uparrow}$ and $\varphi_\text{1}^{\downarrow}$) levels in the band gap, so there are only the defect states associated with the dangling bond of Si2 ($\varphi_\text{2}^{\uparrow}$ and $\varphi_\text{2}^{\downarrow}$), as shown in the density of states in \textbf{Figure~\ref{Fig4}d}. Meanwhile, since the O1 atom forms a bond with the Si1 atom and becomes threefold coordinated, three Si-O bonds around the O1 atom are elongated, reducing the hybridization between Si s orbital and O s orbital. Consequently, the corresponding antibonding ($\varphi_\text{Si-O}^{*\uparrow}$ and $\varphi_\text{Si-O}^{*\uparrow}$) levels are shifted downward. On some O sites, the hybridization is significantly weakened, so the antibonding levels drop below the CBM level and act as defect-state levels in the band gap (\textbf{Figure~\ref{Fig4}d}); while on other sites, the hybridization weakening is not strong and the levels are still above the CBM level. \textbf{(ii)} For the right hole case, the captured hole is trapped to the right Si2 atom, meaning that the Si2 dangling-bond electron occupying the $\varphi_\text{2}^{\uparrow}$ level of the neutral left-back-projected configuration (\textbf{Figure~\ref{Fig2}d}) is taken away, so the Si2 atom may displace towards the right side. For some V$_\text{O}$ sites, the displaced Si2 atom stays on the O plane determined by its three neighboring O atoms and forms shorter Si-O bonds with the three O atoms. After the large displacement of Si2, the associated $\varphi_\text{2}^{\uparrow}$ and $\varphi_\text{2}^{\downarrow}$ defect states are changed, and now four defect states around the displaced Si2 introduce four levels ($\varphi_\text{2,s}^{\uparrow}$, $\varphi_\text{2,s}^{\downarrow}$, $\varphi_\text{2,p}^{\uparrow}$, $\varphi_\text{2,p}^{\downarrow}$) in the band gap, as shown in \textbf{Figure~\ref{Fig4}d}. The $\varphi_\text{2,s}^{\uparrow}$ and $\varphi_\text{2,s}^{\downarrow}$ states are contributed mainly by the anti-bonding component of Si 3s and O 2s and 2p hybridization; $\varphi_\text{2,p}^{\uparrow}$ and $\varphi_\text{2,p}^{\downarrow}$ states are contributed mainly by the anti-bonding component of Si 3p and O 2p hybridization, and their p-like wavefunctions are obviously perpendicular to the O plane. For some V$_\text{O}$ sites, the displacement of Si2 atom is so large that it goes through the O plane, forming a bond with another O (O1) atom on the right side and becoming fourfold coordinated, or forming two bonds with both an O atom (O1) and a Si atom (Si3) on the right side and becoming fivefold coordinated, as shown in Figure S13. These changes lead to the transition from left-back-projected configuration to double-back-projected configuration. In these +1 charged double-back-projected configurations, the $\varphi_\text{2}^{\uparrow}$ and $\varphi_\text{2}^{\downarrow}$ states are eliminated due to the passivation of Si2 dangling bond, but an antibonding state ($\varphi_\text{Si-O}^{*\uparrow}$ and $\varphi_\text{Si-O}^{*\downarrow}$) drops from the conduction band into the band gap because of the elongated Si-O bond on the right side, which gives a density of states similar to that of double-back-projected configuration in \textbf{Figure~\ref{Fig4}d}.

The structure of the in-plane configuration can be considered as an intermediate state on the transition from Si-dimer to back-projected configuration, as shown by the right-in-plane configuration (with a hole trapped on the right-side Si2 atom) in \textbf{Figure~\ref{Fig4}b}. In the Si-dimer configuration of V$_\text{O}$, the Si-Si dimer stays near the vacancy site, while in the right-back-projected configuration, the Si2 atom on the right side moves away from the vacancy site, goes through the O plane and stays on the right side of the O plane. If the Si2 atom just stays on the O plane, it is called the right-in-plane configuration. Similar to the asymmetry found for left- and right-back-projected configurations on one O site, there are also left-in-plane (the hole is trapped on the left-side Si1) and right-in-plane (hole on right-side Si2) configurations, which are both classified as in-plane configurations.  For +1 charged V$_\text{O}$ on some O sites, the Si-dimer configuration cannot be stable during structural relaxation and transforms to the in-plane configurations spontaneously, because the two Si cations around the +1 charged V$_\text{O}$ repel each other and break the Si-Si dimer bond during structural relaxation. For neutral V$_\text{O}$ on the same O site, the Si-dimer configuration is always stable and does not transform to the in-plane configurations during structural relaxation. Even if the initial structure is perturbed intentionally to the in-plane configuration, the structure cannot be maintained and will relax to the Si-dimer or back-projected configurations for neutral V$_\text{O}$. The electronic structure of right-in-plane configuration is similar to that of left-back-projected (right hole) configurations. As shown in \textbf{Figure~\ref{Fig4}d}, there exist 6 defect levels in the band gap of the right-in-plane configuration. The dangling bond of Si1 in \textbf{Figure~\ref{Fig4}b} creates two defect-state levels, the spin-up $\varphi_\text{1}^\uparrow$ level is low and occupied, while the spin-down $\varphi_\text{1}^\downarrow$ level is high due to the exchange splitting and unoccupied. The Si2 atom with the trapped hole produces four states ($\varphi_\text{2,s}^\uparrow$, $\varphi_\text{2,s}^\downarrow$, $\varphi_\text{2,p}^\uparrow$, $\varphi_\text{2,p}^\downarrow$), which are contributed also by the anti-bonding components of the hybridization between Si2 and three O atoms on the plane.

The twisted configuration is characterized by the twisting of Si-O bonds, as shown in \textbf{Figure~\ref{Fig4}c}. The O1 atom adjacent to Si1 atom is twisted from the original site on the left side of Si1 to the O vacancy site on the right side of Si1, reconnecting the two Si atoms through the new Si1-O1-Si2 bonds. This was also called as the forward-oriented configuration.\cite{RN7} After this process, the right-side Si2 atom becomes coordinated by four O atoms, while the left-side Si1 atom is still coordinated by three O atoms. Therefore, the dangling bond at the Si2 atom is passivated and the related defect states are eliminated in the band gap. This is shown by the density of states of the twisted configuration in \textbf{Figure~\ref{Fig4}d}, where only two defect states $\varphi_\text{1}^\uparrow$ and $\varphi_\text{1}^\downarrow$ appear in the band gap. Meanwhile, the Si-O bonds around the threefold O1 are significantly elongated during the twisting distortion, so the corresponding antibonding levels $\varphi_\text{Si-O}^{*\uparrow}$ and $\varphi_\text{Si-O}^{*\downarrow}$ are shifted downward in energy and stay near the CBM level due to the weakened hybridization, as shown in \textbf{Figure~\ref{Fig4}d}.

After identifying all the possible configurations of positively charged V$_\text{O}$ on 5 O sites, we extend the structural search for the remaining 139 O sites. We suppose that V$_\text{O}$ on these sites can also take the seven identified configurations, so the initial structure of V$_\text{O}$ is perturbed to match the seven configurations for each O site. Structural relaxation of these configurations shows that the relaxed structures fall into the scope of the seven identified configurations, and no new structure is found for these perturbed initial structures of V$_\text{O}$. 

In \textbf{Figure~\ref{Fig4}e, f}, the distribution of the formation energies of those configurations in +1 charge state are plotted. Different from the energy distribution of neutral V$_\text{O}$, the energies of +1 charged V$_\text{O}$ are much more complicated. \textit{Firstly}, the formation energy of Si-dimer configuration ranges from 6.52 eV to 8.36 eV, which is no longer apparently lower than the those of other configurations. \textit{Secondly}, compared to the Gaussian-like distributions of neutral V$_\text{O}$ formation energy, the distributions of +1 charged V$_\text{O}$ formation energy in \textbf{Figure~\ref{Fig4}e, f} are more uniform and have a wider energy range. For instance, the neutral back-projected V$_\text{O}$ has a range of about 2 eV, but the range of +1 charged back-projected V$_\text{O}$ increases to about 3 eV. \textit{Thirdly}, the in-plane and twisted configurations of +1 charged V$_\text{O}$ can have quite low formation energies. The lower limit of the formation energy of in-plane configuration is 5.67 eV, and that of twisted configuration is 6.27 eV, both lower than that of Si-dimer configuration, showing these two configurations can indeed exist in a-SiO$_\text{2}$.

Depending on different O sites, all the 7 configurations have a possibility to act as the ground-state structure of +1 charged V$_\text{O}$, and the possibilities are all non-negligible, as shown in \textbf{Figure~\ref{Fig4}g}. The proportion of Si-dimer configuration as the ground state can be as high as 97.2\% for neutral V$_\text{O}$ (\textbf{Figure~\ref{Fig3}a}), but the proportion is significantly reduced to 35.4\% for +1 charged V$_\text{O}$. On the contrary, the proportion of back-projected configuration for neutral V$_\text{O}$ (\textbf{Figure~\ref{Fig3}a}) is only 2.8\%, but it is elevated to 20.1\% for +1 charged V$_\text{O}$. The proportions of in-plane, double-back-projected, and twisted configurations are 20.1\%, 14.6\%, and 9.7\%, respectively. It should be noted that their possibility in acting as the ground-state structure of +1 charged V$_\text{O}$ has never been reported in literature. Our result is in contrast to the previously calculated results \cite{RN11}, which show that only the back-projected configuration is the ground-state structure of +1 charged V$_\text{O}$. This is actually the fundamental requirement of the four-state model. Recent first-principles calculation studies also reported that the Si-dimer configuration is the ground-state structure of +1 charged V$_\text{O}$ in a-SiO$_\text{2}$ \cite{RN18}, which is also in contrast with our results. The misleading findings in these previous studies should stem from their insufficient consideration of all possible structural configurations, i.e., only a limited number of perturbations were added to the V$_\text{O}$ structures, so the structures all relax to the Si-dimer or back-projected configurations during the following local structural relaxation. Since each configuration has a non-negligible possibility in acting as the ground-state structure, the future studies on +1 charged V$_\text{O}$ defects in a-SiO$_\text{2}$ should consider all these configurations and the ignorance of a certain configuration may cause misunderstanding in the microscopic mechanisms. 

\section{\label{sec:level1}Discussion}

\subsection{\label{sec:level2}From four-state model to all-state model}

According to the calculated results above, neutral V$_\text{O}$ has four configurations, three of which can act as the ground state; +1 charged V$_\text{O}$ has seven configurations, and each of them can act as the ground state, as shown in \textbf{Figure~\ref{Fig4}g}. These results indicate that the ground-state structures of V$_\text{O}$ are complicated and rely on the atomic coordination around the O site. All these behaviors are in stark contrast to that of V$_\text{O}$ in crystalline SiO$_\text{2}$. Given these results, we will now discuss whether the complicated ground-state structures support the four-state model and how these configurations of V$_\text{O}$ found in a-SiO$_\text{2}$ influence the NBTI degradation of Si/SiO$_\text{2}$ pMOSFET. 

\begin{figure*}[htbp]
\centering
\includegraphics[width=0.6\textwidth]{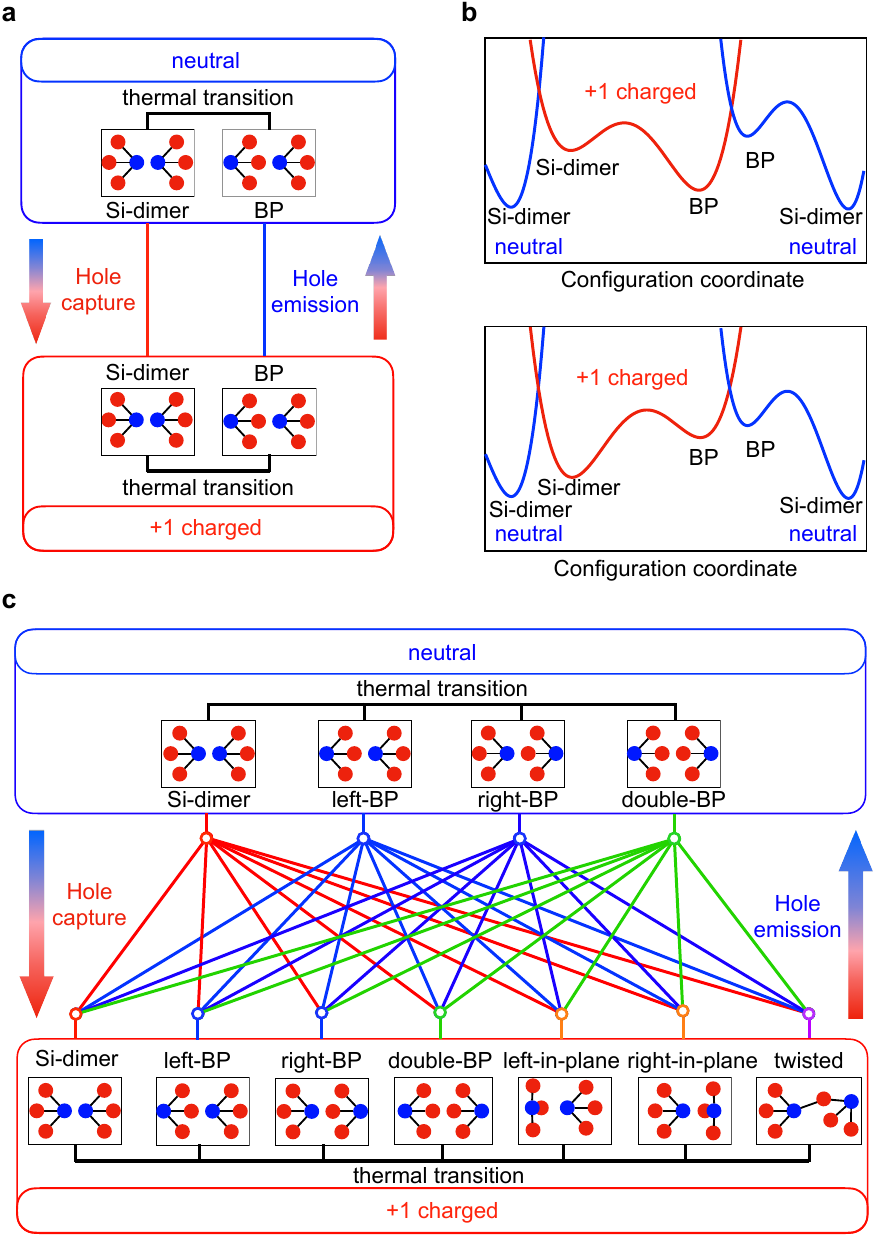}
\caption{\textbf{NMP charge-state transitions during hole capture/emission and thermal transitions of V$_\text{O}$ defects considered in the four-state model and our all-state model.} (a) Two NMP charge-state transitions and two thermal transitions considered in the four-state model. (b) Potential energy surface of V$_\text{O}$ required by the four-state model (upper panel), and an example of V$_\text{O}$ potential energy surface that does not support the four-state model (lower panel). (c) 28 NMP charge-state transitions, 6 thermal transitions between the neutral V$_\text{O}$ configurations, and 21 thermal transitions between the +1 charged V$_\text{O}$ configurations considered in our all-state model.}\label{Fig5}
\end{figure*}

As illustrated in \textbf{Figure~\ref{Fig1}a} and \textbf{Figure~\ref{Fig5}a}, when pMOSFET is under NBTI stress (negative and large V$_\text{G}$), the V$_\text{O}$ defect level in a-SiO$_\text{2}$ is shifted up relative to the VBM level of Si channel, and thus neutral V$_\text{O}$ can capture holes from Si channel, becoming +1 charged. With the cumulation of the positively charged V$_\text{O}$ in a-SiO$_\text{2}$ layer, a more negative V$_\text{th}$ is required to turn on the pMOSFET, giving a negative $\Delta$V$_\text{th}$ under NBTI stress. When V$_\text{G}$ is removed or decreased to a negative but small value (the pMOSFET is under recovery), the V$_\text{O}$ level is shifted down and the holes captured by +1 charged V$_\text{O}$ will be emitted backward to Si channel. Therefore, V$_\text{O}$ in a-SiO$_\text{2}$ becomes neutral again, which gives a less negative V$_\text{th}$ and $\Delta$V$_\text{th}$ (a small absolute value of $\Delta$V$_\text{th}$). During the recovery, the time-dependent defect spectroscopy (TDDS) shows that the absolute value of V$_\text{th}$ drops in discrete steps, occurring at stochastic times \cite{RN54}. Each discrete step is attributed to the emission of a single trapped hole, so each defect has unambiguous fingerprints, emission time constants ($\tau_e$) and step heights, which can be extracted from TDDS measurements \cite{RN54}. Waltl et al. used different V$_\text{G}$ during recovery to study the influence of recovery V$_\text{G}$ on $\tau_e$ of defects. They observed two types of defects with different $\tau_e$ dependence on recovery V$_\text{G}$, i.e., for the first type (fixed trap), $\tau_e$ is insensitive to the recovery V$_\text{G}$, while for the other type (switching trap), $\tau_e$ becomes shorter if the recovery V$_\text{G}$ is less negative (the larger the V$_\text{G}$ change from stress to recovery is, the faster the defect emits holes) \cite{RN69,RN64}. As discussed above, the emission of holes from +1 charged V$_\text{O}$ to Si channel makes V$_\text{th}$ shift to a less negative value during the recovery. Therefore, when the recovery V$_\text{G}$ becomes less negative, the V$_\text{O}$ defect level is shifted down to a lower position, and then the emission of holes from the lower V$_\text{O}$ level to the higher Si VBM level becomes faster, giving rise to a shorter $\tau_e$. This means all the defects observed in the TDDS measurement should have $\tau_e$ sensitive to the recovery V$_\text{G}$. However, only the behavior of switching traps follows this expectation, while the behavior of fixed traps ($\tau_e$ insensitive to the recovery V$_\text{G}$) cannot be explained. To understand the puzzling $\tau_e$ behavior of fixed traps and also other abnormal observations (see Ref. \cite{RN54}) about the capture time constant of defects, Grasser et al. proposed the four-state model. In the explanation to NBTI recovery behavior above, the V$_\text{O}$ defect is supposed to have only two states, the ground state of neutral V$_\text{O}$ and the ground state of +1 charged V$_\text{O}$, so it was also called as the two-state model \cite{RN14}. However, in the four-state model, they pointed out that both neutral and +1 charged V$_\text{O}$ have two structural configurations, i.e., neutral V$_\text{O}$ not only has the ground-state Si-dimer configuration but also has a metastable back-projected (BP) configuration; +1 charged V$_\text{O}$ not only has the ground-state BP configuration but also has a metastable Si-dimer configuration \cite{RN12}. According to the structural transition diagram in \textbf{Figure~\ref{Fig5}a}, the hole emission during recovery has two different pathways: BP (q=+1) $\rightarrow$ BP (q=0) $\rightarrow$ Si-dimer (q=0), and BP (q=+1) $\rightarrow$ Si-dimer (q=+1) $\rightarrow$ Si-dimer (q=0).  For the first pathway, the emission time constant $\tau_e$ is contributed mainly by the BP (q=+1) $\rightarrow$ BP (q=0) hole emission step, while the following BP (q=0) $\rightarrow$ Si-dimer (q=0) thermal transition step is quick and the time can be neglected because the transition is from the metastable back-projected configuration to the ground-state Si-dimer configuration, as shown in the upper panel of \textbf{Figure~\ref{Fig5}b}. Since the transition barrier and thus the time of the BP (q=+1) $\rightarrow$ BP (q=0) hole emission step can be changed by changing V$_\text{G}$, $\tau_e$ of the first pathway depends on V$_\text{G}$, which can explain the behavior of switching traps. For the second pathway, $\tau_e$ is contributed mainly by the BP (q=+1) $\rightarrow$ Si-dimer (q=+1) thermal transition step, while the Si-dimer (q=+1) $\rightarrow$ Si-dimer (q=0) hole emission step is quick. Since the transition barrier and thus the time of the BP (q=+1) $\rightarrow$ Si-dimer (q=+1) thermal transition step is insensitive to V$_\text{G}$, $\tau_e$ of the second pathway is insensitive to V$_\text{G}$. In this way, the four-state model can also explain the puzzling behavior of the fixed traps \cite{RN64}. From the two-state model to the four-state model, the introduction of the metastable configurations of V$_\text{O}$ in each charge state explains successfully the behaviors of both fixed and switching traps, therefore the four-state model has attracted wide attention and becomes the most important model in BTI-related reliability physics studies in MOSFET \cite{RN59,RN60,RN61,RN62,RN63}.

However, our first-principles calculations in \textbf{Figure~\ref{Fig3}} and \textbf{Figure~\ref{Fig4}} show clearly that both neutral and +1 charged V$_\text{O}$ on different O sites in a-SiO$_\text{2}$ have multiple ground-state and metastable configurations, which are not considered in the four-state model unfortunately. In order to include the roles of all possible V$_\text{O}$ configurations in the NBTI stress and recovery processes, we propose an all-state model that considers the complicated variance of the ground-state and metastable V$_\text{O}$ configurations on the non-equivalent O sites in a-SiO$_\text{2}$. For neutral V$_\text{O}$, the four-state model only considers one metastable configuration (back-projected), but the energy of the double-back-projected configuration can be as low as that of the back-projected configuration (\textbf{Figure~\ref{Fig3}d}), so its role in NBTI cannot be neglected as well. In our all-state model, the metastable configurations include not only the back-projected (left and right back-projected) configurations but also the double-back-projected configuration. On the other hand, our calculations showed that the ground state of neutral V$_\text{O}$ is not necessarily the Si-dimer configuration, i.e., the Si-dimer configuration is the ground state on 97.2\% of O sites, but there are still 2.8\% of O sites on which the back-projected configurations become the ground state, as shown in \textbf{Figure~\ref{Fig3}c}. The four-state model considers only the Si-dimer configuration as the ground state of neutral V$_\text{O}$, which is also questionable, so our all-state model considers both situations. For +1 charged V$_\text{O}$, besides the Si-dimer, back-projected and double-back-projected configurations, there are also in-plane (left- and right-in-plane) and twisted configurations, and the back-projected configurations are the ground state only for 20.1\% O sites, while the Si-dimer, double-back-projected, in-plane, twisted configurations can be the ground state for 35.4\%, 14.6\%, 20.1\%, 9.7\% O sites, respectively. However, the four-state model only considers the back-projected configuration as the ground state and the Si-dimer configuration as the metastable state, which neglects obviously many other possible ground-state and metastable configurations and is thus valid only for a small proportion of V$_\text{O}$ defects. Our all-state model considers the non-equivalency of all O sites in a-SiO$_\text{2}$, i.e., on each O site, the ground-state configuration and the corresponding metastable configurations are determined based on the first-principles energy calculations, and these configurations vary on the different O sites. In this way, all the configurations in different charge states are completely included in the all-state model.

Besides including all the V$_\text{O}$ states (structural configurations in different charge states), the all-state model also includes all the transitions between these states. Depending on whether the charge state is changed, the transitions are classified into two types. When V$_\text{O}$ captures or emits a hole, the defect charge state is changed between 0 and +1, such a transition is a nonradiative multiphonon (NMP) transition, as shown by the red, blue and green lines in \textbf{Figure~\ref{Fig5}c}. When V$_\text{O}$ stays in the same charge state and changes its structural configuration, the transition is a pure thermal transition, as shown by the horizontal black lines in \textbf{Figure~\ref{Fig5}c}. In the following, we will show how the all-state model considers all these transitions.

For the NMP transitions with hole capture and emission, they can occur between different structural configurations of neutral and +1 charged V$_\text{O}$. For example, during NBTI recovery, the hole emission can make the back-projected configuration of +1 charged V$_\text{O}$ transit into not only the back-projected configuration, but also the Si-dimer and double-back-projected configurations of neutral V$_\text{O}$. However, the four-state model only considers the transition between the back-projected configurations (\textbf{Figure~\ref{Fig5}a}), while neglects the transition to the Si-dimer and other configurations, because the NMP transition with large structural change (e.g., from back-projected to Si-dimer configuration) was assumed to be impossible \cite{RN54}. However, this assumption may be unreasonable. Recent studies demonstrated that the defect-induced NMP transition with large structural change is important and cannot be neglected \cite{RN66,RN68}. For instance, the hole capture transition induced by $\text{Ga}_\text{Cu}$ in CuGaSe$_\text{2}$ is very fast, although it is accompanied by a large structural change, as reflected by a large configuration coordinate difference ($\Delta$Q $\approx$ 15 amu$^\text{1/2}$ \AA) \cite{RN67}. According to the definition of Huang-Rhys factor S = 0.5$\omega$($\Delta$Q)$^\text{2}$/$\hbar$ \cite{RN65}, the NMP transitions with large $\Delta$Q have stronger electron-phonon couplings and can have a high transition rate. Therefore, the NMP transitions with large structural change should not be neglected. Our calculations show that the Si-dimer-$\textit{to}$-Si-dimer transition from neutral to +1 charged V$_\text{O}$ has a $\Delta$Q around 3.5 amu$^\text{1/2}$ \AA, while the Si-dimer-$\textit{to}$-BP, Si-dimer-$\textit{to}$-double-BP, Si-dimer-$\textit{to}$-in-plane, Si-dimer-$\textit{to}$-twisted transitions have much larger $\Delta$Q around 10-20 amu$^\text{1/2}$ \AA (as shown in Figure S14), so these transitions may have large transition rates and thus should be considered. In the all-state model, all the NMP transitions regardless of $\Delta$Q are explicitly considered. For instance, hole emission transition can occur from one of 7 possible +1 charged V$_\text{O}$ configurations to one of 4 possible neutral V$_\text{O}$ configurations. As shown by the red, blue and green lines plotted in \textbf{Figure~\ref{Fig5}c}, there are totally 28 structural transition pathways for hole capture and hole emission, respectively.

For the thermal transition between different configurations in the same charge state, our all-state model also includes all the possible transitions. Since the thermal transition barriers are between different configurations in the same charge state and are thus irrelevant to V$_\text{G}$ in MOSFET devices (as shown in \textbf{Figure~\ref{Fig5}b}), its transition rate and time are also independent of V$_\text{G}$. This character makes the introduction of the thermal transition in the four-state model explain successfully the V$_\text{G}$-independent emission time constants of the traps observed by TDDS, which demonstrates the importance of thermal transition in NBTI. However, the four-state model only considers one thermal transition for +1 charged V$_\text{O}$ (from back-projected to Si-dimer configuration), while the back-projected configuration can also transit into the double-back-projected, in-plane and twisted configurations of +1 charged V$_\text{O}$.  Therefore, in our all-state model, we consider all the 6 thermal transitions from back-projected (left or right) configuration to 6 other configurations of +1 charged V$_\text{O}$. Since there are in total 7 configurations of +1 charged V$_\text{O}$, the number of possible thermal transitions is as large as 21, as shown in \textbf{Figure~\ref{Fig5}c}. Similarly, there are 6 possible thermal transitions for the 4 configurations of neutral V$_\text{O}$. They are all considered in our all-state model.

As discussed above, from the two-state model to four-state model and our all-state model, the numbers of possible states and transitions considered in the NBTI stress and recovery processes increase, which makes the models more and more comprehensive and reasonable. In the two-state model, only one structural configuration is considered for each charge state (0 and +1), and only one NMP transition is considered for hole capture and emission; while in the four-state model, two configurations (one ground state and one metastable state) are considered for each charge state, and 2 NMP transitions and 2 thermal transitions are considered for hole capture and emission. Here, according to our systematical first-principles calculations, there are 4 configurations of neutral V$_\text{O}$ and 7 configurations of +1 charged V$_\text{O}$, so there should be 28 NMP transitions during hole capture and emission, 6 thermal transitions between the neutral V$_\text{O}$ configurations, and 21 thermal transitions between the +1 charged V$_\text{O}$ configurations. As a result, we propose the all-state model which considers all these configurations and all these possible transitions. We believe the advancement from the four-state model to the all-state model may provide more comprehensive understandings for MOSFET reliability physics.

\subsection{\label{sec:level2}Impact of all-state model on NBTI reliability physics: importance of V$_\text{O}$ defects}

One important impact on the Si/SiO$_\text{2}$ MOSFET reliability physics that our all-state model has is about the origin defect of NBTI, which is in sharp contrast with the origin defect proposed recently based on the four-state model. In the initial study of the four-state model, the 0/+1 charge-state transitions and thermal transitions among the four states of the V$_\text{O}$ defect were used to explain NBTI and abnormal behavior of emission time constants during NBTI recovery. However, in the following studies, the authors from the same group pointed out \cite{RN64} that “The chemical nature of the BTI defect is still controversial \cite{RN74,RN75,RN76} and many possible defect candidates have been suggested over the years. For NBTI the oxygen vacancy must be discarded as a possible hole trap due a too low trap level \cite{RN76}.” The reason they excluded V$_\text{O}$ as the origin defect of NBTI is that V$_\text{O}$ with very low 0/+1 trap levels cannot capture holes and play roles in NBTI. With the negative V$_\text{G}$, the V$_\text{O}$ trap level in a-SiO$_\text{2}$ is shifted upward relative to the Si VBM level. Only when the upward-shifted V$_\text{O}$ level is higher than the Fermi level of Si (near VBM level in pMOSFET), can the V$_\text{O}$ defect capture holes from the Si channel. The operating V$_\text{G}$ of Si/SiO$_\text{2}$ pMOSFET is usually around $-1$ V \cite{RN79}. If the V$_\text{O}$ trap levels are very low (1 eV below the Si VBM), they are still below the Si VBM level even after they are shifted upward by the negative V$_\text{G}$, and thus they cannot capture holes and play roles in NBTI. 

\begin{figure*}[htbp]
\centering
\includegraphics[width=0.8\textwidth]{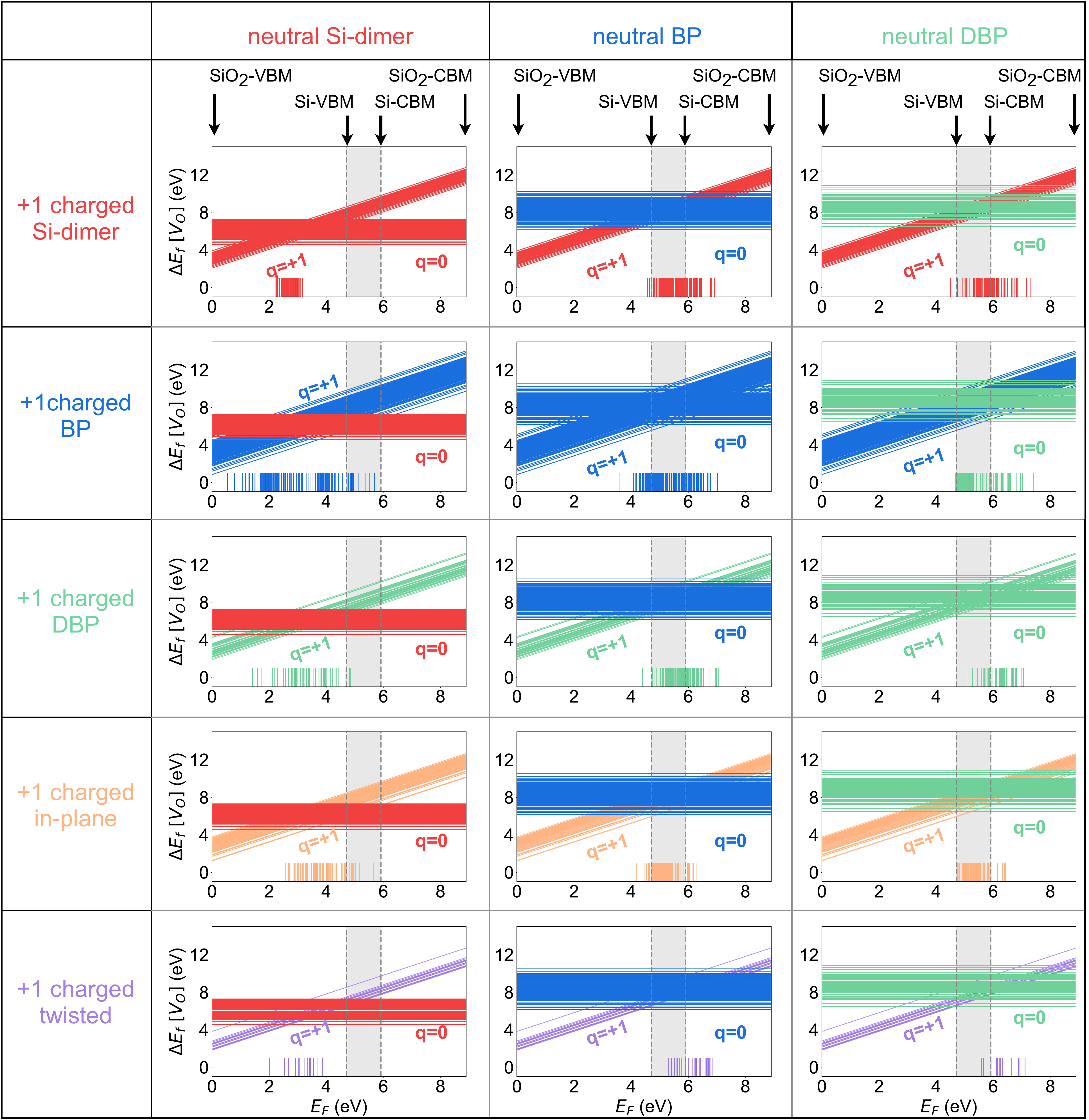}
\caption{\textbf{The formation energy and (0/+1) transition level distribution of V$_\text{O}$ on different O sites in a-SiO$_\text{2}$.} Each panel shows the formation energies of one neutral V$_\text{O}$ configuration (q=0, horizontal lines) and one +1 charged V$_\text{O}$ configuration (q=+1, oblique lines), as functions of Fermi level. For the q=0 and q=+1 configurations, they both have dozens of lines which show their formation energies on different O sites (only when the configuration is stable, as a ground state or metastable state, its formation energy on the site can be calculated. Since there are 144 O sites, the maximum number of lines is 144). For a certain O site, when both the neutral and +1 charged configurations are stable, the (0/+1) transition level between them is plotted as the short vertical line on the $E_F$ axis. The VBM and CBM levels of a-SiO$_\text{2}$ are located at $E_F$=0 and $E_F$=8.9 eV, respectively. The dashed vertical lines at $E_F$=4.71 eV and $E_F$=5.91 eV show the VBM and CBM levels of Si. The five panels in the first column show the transitions from the neutral Si-dimer configuration to the +1 charged Si-dimer, back-projected (BP), double-back-projected (DBP), in-plane, twisted configurations. The second (third) column shows the transitions from the neutral BP (neutral DBP) configuration to the five +1 charged configurations.}\label{Fig6}
\end{figure*}

According to our calculations, neutral V$_\text{O}$ has 4 configurations and +1 charged V$_\text{O}$ has 7 configurations, and different configurations can have very different formation energies, so there are 28 types of (0/+1) charge-state transitions with the levels scattered in the a-SiO$_\text{2}$ band gap. Even for the same type of transition, the (0/+1) levels of V$_\text{O}$ on different O sites are still scattered in a wide range due to the non-equivalency of O sites in a-SiO$_\text{2}$. It is questionable whether all these scattered (0/+1) levels are very low. If some of the levels are higher enough for capturing holes, discarding V$_\text{O}$ as an origin defect of NBTI can be misleading. In order to have a comprehensive account of the V$_\text{O}$ (0/+1) transition levels, we show in \textbf{Figure~\ref{Fig6}} all the possible (0/+1) levels with different configurations.

When V$_\text{O}$ takes the Si-dimer configuration in both the neutral and +1 charge states, the (0/+1) transition levels are indeed very low, from 2.47 to 1.55 eV below the Si VBM (see the first panel in the first column of \textbf{Figure~\ref{Fig6}}), which is consistent with the results reported in Ref. \cite{RN28}. If only the Si-dimer configurations are considered during hole capture and emission processes, V$_\text{O}$ can indeed be discarded due to the low-lying trap levels \cite{RN64}. However, if we consider the transitions from the Si-dimer configuration of neutral V$_\text{O}$ to other configurations of +1 charged V$_\text{O}$, such as back-projected, double-back-projected and in-plane configurations, the (0/+1) transition levels can be scattered in a wider range, as shown in the 2$^\text{nd}$ to 4$^\text{th}$ panels in the left column of \textbf{Figure~\ref{Fig6}}. On many O sites, the V$_\text{O}$ levels fall in the energy range that is high enough and suitable for hole capture during NBTI stress and hole emission during NBTI recovery of Si/SiO$_\text{2}$ pMOSFET, i.e, from 1 eV below the VBM level of Si to the mid-gap level of Si. As discussed above, if the levels are lower than this range, they cannot capture holes under V$_\text{G}$ = $-1$ V \cite{RN79}. On the other hand, if the levels are too high, e.g., much higher than the Fermi level of p-type Si in pMOSFET, these V$_\text{O}$ defects are always +1 charged, thus they neither capture holes during NBTI stress nor emit holes during NBTI recovery. These high-level V$_\text{O}$ defects do not play roles in NBTI. For the V$_\text{O}$ defects whose levels fall in the energy range required for hole capture and emission, they can play important roles in NBTI and thus should not be ignored. Besides the transitions from the Si-dimer configuration of neutral V$_\text{O}$, there are also transitions from other configurations of neutral V$_\text{O}$, such as the back-projected and double-back-projected configurations of neutral V$_\text{O}$, to different configurations of +1 charged V$_\text{O}$. As shown in the panels in the middle and right columns of \textbf{Figure~\ref{Fig6}}, the (0/+1) levels of these transitions may also fall in the energy range required for hole capture and emission. For example, 4 transitions from the back-projected configuration of neutral V$_\text{O}$ to the dimer, back-projected, double-back-projected and in-plane configurations of +1 charged V$_\text{O}$ have many levels distributed in the range from $-$0.5 eV below the Si VBM level to 0.5 eV above the Si VBM level (see the middle column of \textbf{Figure~\ref{Fig6}}); 3 transitions from the double-back-projected configuration of neutral V$_\text{O}$ to the Si-dimer, back-projected, in-plane configurations of +1 charged V$_\text{O}$ have (0/+1) levels distributed in the range from the Si VBM level to 0.5 eV above the Si VBM level. The V$_\text{O}$ defects with these configurations and transitions can also play important roles in NBTI and thus should not be ignored either.

Consequently, according to the wide energy range of the calculated (0/+1) levels in \textbf{Figure~\ref{Fig6}}, the V$_\text{O}$ defects with different structural configurations on different O sites in a-SiO$_\text{2}$ cannot be discarded as the origin of NBTI. In the recent studies that discarded V$_\text{O}$ as a possible hole trap in NBTI, they neglected the diversity of structural configurations and (0/+) transition levels of V$_\text{O}$ on different O sites in a-SiO$_\text{2}$, which causes them to ignore the role of V$_\text{O}$ in NBTI. Since our all-state model is just proposed to describe the diversity of structural configurations and properties of V$_\text{O}$ in a-SiO$_\text{2}$ explicitly, it becomes obvious for us to notice the importance of V$_\text{O}$, the most famous defect in a-SiO$_\text{2}$, in NBTI. 

Recently other defects in a-SiO$_\text{2}$, such as hydrogen bridge and hydroxyl-$E^\prime$ center \cite{RN12}, were also proposed as possible origins of NBTI. It should be noted that these defects on non-equivalent sites also have various structural configurations and thus many transitions between the configurations in the different charge states, so the all-state model considering all their states and transitions is also required for describing the roles of these defects in NBTI. Furthermore, other reliability issues, such as positive bias temperature instability (PBTI), random telegraph noise (RTN) and stress-induced leakage current (SILC), are also correlated with the carrier capture and emission by defects in the amorphous oxide dielectric layers (such as SiO$_\text{2}$ or HfO$_\text{2}$), so the diversity of structural configurations and transitions of different origin defects should also influence the microscopic mechanisms of these reliability issues. Therefore, our all-state model should be generally adopted for the understanding and modeling of these reliability issues caused by defects in amorphous oxides.

In summary, by comprehensively studying the structural configurations of neutral and +1 charged V$_\text{O}$ defects in amorphous SiO$_\text{2}$, we find neutral V$_\text{O}$ can have four types of configurations and +1 charged V$_\text{O}$ can have seven types of configurations. Among them, three configurations can be the ground-state structures of neutral V$_\text{O}$ and all the seven configurations can be the ground-state structures of +1 charged V$_\text{O}$. Correspondingly, there exist 28 NMP transitions during hole capture and emission, 6 thermal transitions between the configurations of neutral V$_\text{O}$, and 21 thermal transitions between the configurations of +1 charged V$_\text{O}$. The structural diversity and various transitions of V$_\text{O}$ defects cannot be described by the simple bi-stability assumption of the four-state model, which makes the four-state model invalid for describing the reliability physics induced by oxide defects accurately.  Therefore, we propose an all-state model to describe all the structural configurations and transitions of defects in dielectric oxides, so that their roles in NBTI and other reliability issues of Si/SiO$_\text{2}$ MOSFETs can be accurately modelled. With the all-state model, we re-evaluated the role of V$_\text{O}$ in NBTI of Si/SiO$_\text{2}$ pMOSFETs and found that the large number of V$_\text{O}$ (0/+1) transition levels in a wide energy range make them effective hole trap centers in a-SiO$_\text{2}$, so they can be an origin defect of NBTI, which challenges the recent studies that discarded V$_\text{O}$. Given that the dielectric layers in many MOSFET (including FinFET and GAAFET) devices all have amorphous structures and thus their defects also have similar structural diversity and complicated transitions, the all-state model that we proposed should be general for modeling dielectric-defect-induced BTI, RTN, SILC and other reliability issues of all these devices, no matter the devices have the Si, Ge, SiC, GaN, MoS$_2$, WSe$_2$ or other new types of semiconductor channels.

\section{\label{sec:level1}Methods}

\subsection{\label{sec:level2}First-principles calculations}

Calculations are performed based on density functional theory as implemented in the Vienna Ab initio Simulation Package (VASP) \cite{RN29}. The projector augmented-wave (PAW) \cite{RN30} pseudopotentials are used. The cutoff energy for the plane-wave basis is set to 400 eV. Hybrid functional of Heyd-Scuseria-Ernzerhof (HSE) \cite{RN31} form with an exchange parameter of 48\% is used to reproduce the 8.9 eV band gap of amorphous SiO$_\text{2}$ in experiments \cite{RN45}. HSE is also adopted for all of the defect-related calculations combined with $\Gamma$-point sampling, including the structural relaxation and total energy calculations of 144 possible V$_\text{O}$ sites in both neutral and +1 charged states, so that the localized nature of each defect state can be ensured. 

\subsection{\label{sec:level2}Generation of a-SiO$_\text{2}$ structures}

To generate a reasonable a-SiO$_\text{2}$ supercell model, we adopt the bond-switching Monte Carlo (MC) method \cite{RN21}, which can effectively mimic the atomic bonding conditions in real amorphous structures. In the bond-switching MC simulation, $\alpha$-quartz SiO$_\text{2}$ is chosen as the initial structure. The valence force field is adopted to relax the structure and achieve the total energy, and then Metropolis algorithm is used to determine whether the relaxed structure is accepted. To test the convergence against the supercell size, we rigorously calculate the inverse participation ratio (IPR) of the highest occupied molecular orbital (HOMO) and the lowest unoccupied molecular orbital (LUMO) respectively in 216-atom and 96-atom supercell structures, and find the HOMO in the 96-atom supercell structure is actually a partially-localized state caused by O-Si-O bond stretching, while that in 216-atom supercell is well delocalized. The localization of HOMO in 96-atom supercell structure is a reflection of defect state, while the delocalization of HOMO in a 216-atom supercell structure properly mimics the character of SiO$_\text{2}$ VBM state, suggesting that using a smaller supercell may be inadequate to get reliable results.

\subsection{\label{sec:level2}Defect simulations}

To calculate the formation energies of V$_\text{O}$ on 144 non-equivalent O sites in the a-SiO$_\text{2}$ supercell, we adopt the well-established defect theory \cite{RN32} as implemented in Defect and Dopant ab-initio Simulation Package (DASP) code \cite{RN20}. The formation energy of V$_\text{O}$ in +1 charge state can be calculated by,
\begin{equation}
    \Delta E_f\left(V_O^{+}\right)=E_{t o t}\left(V_O^{+}\right)-E_{tot}(host)+\mu_O+E_F+E_{corr },
\end{equation}
where $E_{t o t}\left(V_O^{+}\right)$ and $E_{tot} (host)$ are the total energies of the amorphous supercell with and without a V$_\text{O}$ defect, and $\mu_O$ is the chemical potential of oxygen. $E_F$ is the Fermi level referenced to the valence band maximum of host supercell. $E_{corr}$ accounts for the spurious interaction in charged defect calculation caused by limited supercell size \cite{RN33}. In the Supplementary Information, the correction energy for each defect is shown.

To search for the possible metastable configurations of V$_\text{O}$ in both neutral and positive charge states, we use the distorted-structure generation scheme implemented in DASP to add structural perturbations for V$_\text{O}$ configurations on all 144 O sites. Mass-weighted structural difference $\Delta$Q is adopted to ensure the perturbations are different from each other,
\begin{equation}
    \Delta Q=\sqrt{\sum_\alpha m_\alpha\left(R_1-R_2\right)^2},
\end{equation}
where $m_\alpha$ is the mass of the $\alpha^\text{th}$ atom in the supercell; $R_1$ and $R_2$ are the cartesian coordinates of each two perturbed structures. Due to the use of HSE functional during all of the atomic structural relaxations for different metastable configurations, the total computational cost is extremely huge, exceeding $10^7$ CPU core hours.

\section{\label{sec:level1}Data availability}

The structures of V$_\text{O}$ defects and their formation energies supporting the key findings of this article are available within the article and the Supplementary Information file. All raw data of first-principles calculations are available from the corresponding authors upon reasonable request.

\section{\label{sec:level1}Code availability}

The codes used to post-process the first-principles raw data are available from the corresponding author upon reasonable request.

\begin{acknowledgments}
This work was supported by National Natural Science Foundation of China (12334005, 12188101, 12174060 and 12404089) and Project of MOE Innovation Platform.
\end{acknowledgments}

\bibliography{apssamp}

\end{document}